
\input epsf
\epsfverbosetrue
\epsfclipon
\magnification=\magstep1
\hsize=5.75truein
\vsize=8.75truein
\baselineskip=12.045truept 
\parindent=2.5em     
\parskip=0pt
\def\hi{\noindent \hangindent=2.5em}
\def\bigskip{\vskip 12.045pt}  
\overfullrule=0pt
\font\twelverm=cmr10 at 12truept  
\twelverm    
\def\title#1{\centerline{\bf #1}}
\def\author#1{\bigskip\centerline{#1}}

\def\sec#1{\bigskip\centerline{#1}\bigskip}
\def\captpar{\dimen0=\hsize
             \advance\dimen0 by -1.0truecm
             \par\parshape 1 0.5truecm \dimen0 \noindent}

\def\ie{{\it i.e.}}
\def\etal{{\it et al.}}
\def\go{
\mathrel{\raise.3ex\hbox{$>$}\mkern-14mu\lower0.6ex\hbox{$\sim$}}
}
\def\lo{
\mathrel{\raise.3ex\hbox{$<$}\mkern-14mu\lower0.6ex\hbox{$\sim$}}
}


\def\etal{{\it et al.}}

\def\kms{{\rm\, km/s}}
\def\kpc{\rm\, kpc }
\def\au{\rm\, AU }

\def\msun { \rm {M_\odot}}

\def\pac{Paczy{\'n}ski}
\def\ie{{\it i.e.}}
\def\that{\hat{t}}
\def\spose#1{\hbox to 0pt{#1\hss}}
\def\simlt{\mathrel{\spose{\lower 3pt\hbox{$\mathchar"218$}}
     \raise 2.0pt\hbox{$\mathchar"13C$}}}
\def\simgt{\mathrel{\spose{\lower 3pt\hbox{$\mathchar"218$}}
     \raise 2.0pt\hbox{$\mathchar"13E$}}}


%
\newbox\hdbox%
\newcount\hdrows%
\newcount\multispancount%
\newcount\ncase%
\newcount\ncols
\newcount\nrows%
\newcount\nspan%
\newcount\ntemp%
\newdimen\hdsize%
\newdimen\newhdsize%
\newdimen\parasize%
\newdimen\spreadwidth%
\newdimen\thicksize%
\newdimen\thinsize%
\newdimen\tablewidth%
\newif\ifcentertables%
\newif\ifendsize%
\newif\iffirstrow%
\newif\iftableinfo%
\newtoks\dbt%
\newtoks\hdtks%
\newtoks\savetks%
\newtoks\tableLETtokens%
\newtoks\tabletokens%
\newtoks\widthspec%
%
%
\immediate\write15{%
CP SMSG GJMSINK TEXTABLE --> TABLE MACROS V. 851121 JOB = \jobname%
}%
%
%
\tableinfotrue%
\catcode`\@=11
%
%
\def\tstrut{\vrule height3.1ex depth1.2ex width0pt}%
\def\and{\char`\&}
\def\tablerule{\noalign{\hrule height\thinsize depth0pt}}%
\thicksize=1.5pt
\thinsize=0.6pt
\def\thickrule{\noalign{\hrule height\thicksize depth0pt}}%
\def\ctr#1{\hfil\ #1\hfil}%
%
%
%
%
\tablewidth=-\maxdimen%
\spreadwidth=-\maxdimen%
\def\tabskipglue{0pt plus 1fil minus 1fil}%
%
%
\centertablestrue%
%
%
%
%
\parasize=4in%
\gdef\ARGS{########}
\gdef\headerARGS{####}
\def\@mpersand{&}
{\catcode`\|=13
\gdef\letbarzero{\let|0}
\gdef\letbartab{\def|{&&}}%
\gdef\letvbbar{\let\vb|}%
}
{\catcode`\&=4
\def\ampskip{&\omit\hfil&}
\catcode`\&=13
\let&0
\xdef\letampskip{\def&{\ampskip}}%
\gdef\letnovbamp{\let\novb&\let\tab&}
}
\def\begintable{
   \begingroup%
   \catcode`\|=13\letbartab\letvbbar%
   \catcode`\&=13\letampskip\letnovbamp%
   \def\multispan##1{
      \omit \mscount##1%
      \multiply\mscount\tw@\advance\mscount\m@ne%
      \loop\ifnum\mscount>\@ne \sp@n\repeat%
   }
   \def\|{%
      &\omit\widevline&%
   }%
   \ruledtable
}
\long\def\ruledtable#1\endtable{%
%
%
%
   \offinterlineskip
   \tabskip 0pt
   \def\widevline{\vrule width\thicksize}
   \def\endrow{\@mpersand\omit\hfil\crnorm\@mpersand}%
   \def\crthick{\@mpersand\crnorm\thickrule\@mpersand}%
   \def\crthickneg##1{\@mpersand\crnorm\thickrule
          \noalign{{\skip0=##1\vskip-\skip0}}\@mpersand}%
   \def\crnorule{\@mpersand\crnorm\@mpersand}%
   \def\crnoruleneg##1{\@mpersand\crnorm
          \noalign{{\skip0=##1\vskip-\skip0}}\@mpersand}%
   \let\nr=\crnorule
   \def\endtable{\@mpersand\crnorm\thickrule}%
   \let\crnorm=\cr
%
%
   \edef\cr{\@mpersand\crnorm\tablerule\@mpersand}%
   \def\crneg##1{\@mpersand\crnorm\tablerule
          \noalign{{\skip0=##1\vskip-\skip0}}\@mpersand}%
   \let\ctneg=\crthickneg
   \let\nrneg=\crnoruleneg
   \the\tableLETtokens
%
%
   \tabletokens={&#1}
%
%
   \countROWS\tabletokens\into\nrows%
   \countCOLS\tabletokens\into\ncols%
%
%
   \advance\ncols by -1%
   \divide\ncols by 2%
   \advance\nrows by 1%
%
%
   \iftableinfo %
      \immediate\write16{[Nrows=\the\nrows, Ncols=\the\ncols]}%
   \fi%
%
%
   \ifcentertables
      \ifhmode \par\fi
      \line{
      \hss
   \else %
      \hbox{%
   \fi
      \vbox{%
         \makePREAMBLE{\the\ncols}
         \edef\next{\preamble}
         \let\preamble=\next
         \makeTABLE{\preamble}{\tabletokens}
      }
      \ifcentertables \hss}\else }\fi
   \endgroup
   \tablewidth=-\maxdimen
   \spreadwidth=-\maxdimen
}
\def\makeTABLE#1#2{
   {
   \let\ifmath0
   \let\header0
   \let\multispan0
%
%
   \ncase=0%
   \ifdim\tablewidth>-\maxdimen \ncase=1\fi%
   \ifdim\spreadwidth>-\maxdimen \ncase=2\fi%
   \relax
%
   \ifcase\ncase %
      \widthspec={}%
   \or %
      \widthspec=\expandafter{\expandafter t\expandafter o%
                 \the\tablewidth}%
   \else %
      \widthspec=\expandafter{\expandafter s\expandafter p\expandafter r%
                 \expandafter e\expandafter a\expandafter d%
                 \the\spreadwidth}%
   \fi %
   \xdef\next{
      \halign\the\widthspec{%
      #1
      \noalign{\hrule height\thicksize depth0pt}
      \the#2\endtable
%
      }
   }
   }
   \next
}
\def\makePREAMBLE#1{
   \ncols=#1
   \begingroup
   \let\ARGS=0
   \edef\xtp{\widevline\ARGS\tabskip\tabskipglue%
   &\ctr{\ARGS}\tstrut}
   \advance\ncols by -1
   \loop
      \ifnum\ncols>0 %
      \advance\ncols by -1%
      \edef\xtp{\xtp&\vrule width\thinsize\ARGS&\ctr{\ARGS}}%
   \repeat
   \xdef\preamble{\xtp&\widevline\ARGS\tabskip0pt%
   \crnorm}
   \endgroup
}
\def\countROWS#1\into#2{
   \let\countREGISTER=#2%
   \countREGISTER=0%
   \expandafter\ROWcount\the#1\endcount%
}%
\def\ROWcount{%
   \afterassignment\subROWcount\let\next= %
}%
\def\subROWcount{%
   \ifx\next\endcount %
      \let\next=\relax%
   \else%
      \ncase=0%
      \ifx\next\cr %
         \global\advance\countREGISTER by 1%
         \ncase=0%
      \fi%
      \ifx\next\endrow %
         \global\advance\countREGISTER by 1%
         \ncase=0%
      \fi%
      \ifx\next\crthick %
         \global\advance\countREGISTER by 1%
         \ncase=0%
      \fi%
      \ifx\next\crnorule %
         \global\advance\countREGISTER by 1%
         \ncase=0%
      \fi%
      \ifx\next\crthickneg %
         \global\advance\countREGISTER by 1%
         \ncase=0%
      \fi%
      \ifx\next\crnoruleneg %
         \global\advance\countREGISTER by 1%
         \ncase=0%
      \fi%
      \ifx\next\crneg %
         \global\advance\countREGISTER by 1%
         \ncase=0%
      \fi%
      \ifx\next\header %
         \ncase=1%
      \fi%
      \relax%
      \ifcase\ncase %
         \let\next\ROWcount%
      \or %
         \let\next\argROWskip%
      \else %
      \fi%
   \fi%
   \next%
}
\def\counthdROWS#1\into#2{%
\dvr{10}%
   \let\countREGISTER=#2%
   \countREGISTER=0%
\dvr{11}%
\dvr{13}%
   \expandafter\hdROWcount\the#1\endcount%
\dvr{12}%
}%
\def\hdROWcount{%
   \afterassignment\subhdROWcount\let\next= %
}%
\def\subhdROWcount{%
   \ifx\next\endcount %
      \let\next=\relax%
   \else%
      \ncase=0%
      \ifx\next\cr %
         \global\advance\countREGISTER by 1%
         \ncase=0%
      \fi%
      \ifx\next\endrow %
         \global\advance\countREGISTER by 1%
         \ncase=0%
      \fi%
      \ifx\next\crthick %
         \global\advance\countREGISTER by 1%
         \ncase=0%
      \fi%
      \ifx\next\crnorule %
         \global\advance\countREGISTER by 1%
         \ncase=0%
      \fi%
      \ifx\next\header %
         \ncase=1%
      \fi%
\relax%
      \ifcase\ncase %
         \let\next\hdROWcount%
      \or%
         \let\next\arghdROWskip%
      \else %
      \fi%
   \fi%
   \next%
}%
{\catcode`\|=13\letbartab
\gdef\countCOLS#1\into#2{%
   \let\countREGISTER=#2%
   \global\countREGISTER=0%
   \global\multispancount=0%
   \global\firstrowtrue
   \expandafter\COLcount\the#1\endcount%
   \global\advance\countREGISTER by 3%
   \global\advance\countREGISTER by -\multispancount
}%
\gdef\COLcount{%
   \afterassignment\subCOLcount\let\next= %
}%
{\catcode`\&=13%
\gdef\subCOLcount{%
   \ifx\next\endcount %
      \let\next=\relax%
   \else%
      \ncase=0%
      \iffirstrow
         \ifx\next& %
            \global\advance\countREGISTER by 2%
            \ncase=0%
         \fi%
         \ifx\next\span %
            \global\advance\countREGISTER by 1%
            \ncase=0%
         \fi%
         \ifx\next| %
            \global\advance\countREGISTER by 2%
            \ncase=0%
         \fi
         \ifx\next\|
            \global\advance\countREGISTER by 2%
            \ncase=0%
         \fi
         \ifx\next\multispan
            \ncase=1%
            \global\advance\multispancount by 1%
         \fi
         \ifx\next\header
            \ncase=2%
         \fi
         \ifx\next\cr       \global\firstrowfalse \fi
         \ifx\next\endrow   \global\firstrowfalse \fi
         \ifx\next\crthick  \global\firstrowfalse \fi
         \ifx\next\crnorule \global\firstrowfalse \fi
         \ifx\next\crnoruleneg \global\firstrowfalse \fi
         \ifx\next\crthickneg  \global\firstrowfalse \fi
         \ifx\next\crneg       \global\firstrowfalse \fi
      \fi
\relax
      \ifcase\ncase %
         \let\next\COLcount%
      \or %
         \let\next\spancount%
      \or %
         \let\next\argCOLskip%
      \else %
      \fi %
   \fi%
   \next%
}%
\gdef\argROWskip#1{%
   \let\next\ROWcount \next%
}
\gdef\arghdROWskip#1{%
   \let\next\ROWcount \next%
}
\gdef\argCOLskip#1{%
   \let\next\COLcount \next%
}
}
}
\def\spancount#1{
   \nspan=#1\multiply\nspan by 2\advance\nspan by -1%
   \global\advance \countREGISTER by \nspan
   \let\next\COLcount \next}%
\def\dvr#1{\relax}%
\def\header#1{%
\dvr{1}{\let\cr=\@mpersand%
\hdtks={#1}%
\counthdROWS\hdtks\into\hdrows%
\advance\hdrows by 1%
\ifnum\hdrows=0 \hdrows=1 \fi%
\dvr{5}\makehdPREAMBLE{\the\hdrows}%
\dvr{6}\getHDdimen{#1}%
{\parindent=0pt\hsize=\hdsize{\let\ifmath0%
\xdef\next{\valign{\headerpreamble #1\crnorm}}}\dvr{7}\next\dvr{8}%
}%
}\dvr{2}}
\def\makehdPREAMBLE#1{
\dvr{3}%
\hdrows=#1
{
\let\headerARGS=0%
\let\cr=\crnorm%
\edef\xtp{\vfil\hfil\hbox{\headerARGS}\hfil\vfil}%
\advance\hdrows by -1
\loop
\ifnum\hdrows>0%
\advance\hdrows by -1%
\edef\xtp{\xtp&\vfil\hfil\hbox{\headerARGS}\hfil\vfil}%
\repeat%
\xdef\headerpreamble{\xtp\crcr}%
}
\dvr{4}}
\def\getHDdimen#1{%
\hdsize=0pt%
\getsize#1\cr\end\cr%
}
\def\getsize#1\cr{%
\endsizefalse\savetks={#1}%
\expandafter\lookend\the\savetks\cr%
\relax \ifendsize \let\next\relax \else%
\setbox\hdbox=\hbox{#1}\newhdsize=1.0\wd\hdbox%
\ifdim\newhdsize>\hdsize \hdsize=\newhdsize \fi%
\let\next\getsize \fi%
\next%
}%
\def\lookend{\afterassignment\sublookend\let\looknext= }%
\def\sublookend{\relax%
\ifx\looknext\cr %
\let\looknext\relax \else %
   \relax
   \ifx\looknext\end \global\endsizetrue \fi%
   \let\looknext=\lookend%
    \fi \looknext%
}%
%
%
\def\tablelet#1{%
   \tableLETtokens=\expandafter{\the\tableLETtokens #1}%
}%
\catcode`\@=12
%

{}~~~
\bigskip
\title{Recent Developments in Gravitational Microlensing and}
\title{the Latest MACHO Results:}
\title{Microlensing Towards the Galactic Bulge}
\bigskip
\centerline {D.P. Bennett$^{\dagger,\ast}$, C.  Alcock$^{\ast,\dagger}$,
R.A.  Allsman$^\ast$, }
\centerline {T.S. Axelrod$^\ast$,
K.H. Cook$^{\ast,\dagger}$, K.C. Freeman$^\ddagger$,}
\centerline {K. Griest$^{\dagger,\|}$, S.L. Marshall$^{\dagger,\flat}$,
S. Perlmutter$^\dagger$,}
\centerline {B.A. Peterson$^\ddagger$, M.R. Pratt$^{\dagger,\flat}$,
P.J. Quinn$^\ddagger$, A.W. Rodgers$^\ddagger$,}
\centerline {C.W. Stubbs$^{\dagger,\clubsuit}$, W. Sutherland$^\spadesuit$}
\centerline { (The MACHO Collaboration) }
\vskip 0.3truein
\centerline{$^\ast$ Lawrence Livermore National Laboratory, Livermore, CA
94550}
\vskip 6pt
\centerline{$^\dagger$ Center for Particle Astrophysics, University of
California,}
\centerline{Berkeley, CA 94720}
\vskip        8pt
\centerline{$^\ddagger$ Mt.  Stromlo and Siding Spring Observatories,}
\centerline{Australian National University, Weston, ACT 2611, Australia}
\vskip 6pt
\centerline{$^\|$ Department of Physics, University of California, \
San Diego, CA 92039 }
\vskip 6pt
\centerline{$^\flat$ Department of Physics, University of California, \
Santa Barbara, CA 93106 }
\vskip 6pt
\centerline{$^\clubsuit$ Department of Astronomy, University of Washington, \
Seattle, WA 98195 }
\vskip 6pt
\centerline{$^\spadesuit$  Department of Physics, University of Oxford,
Oxford OX1 3RH, U.K.}

\sec{ABSTRACT}

We review recent gravitational microlensing results from the EROS, MACHO,
and OGLE collaborations, and present some details of the very latest
MACHO results toward the Galactic Bulge.
The MACHO collaboration has now discovered in excess of 40 microlensing
events toward the Galactic Bulge during the 1993 observing season.
A preliminary analysis of this data suggests a much higher microlensing
optical depth than predicted by standard galactic models suggesting that
these models will have to be revised. This may have important implications for
the structure of the Galaxy and its dark halo.  Also shown are MACHO data of
the first microlensing event ever detected substantially before peak
amplification, the first detection of parallax
effects in a microlensing event, and the first caustic crossing to be
resolved in a microlensing event.

\sec{Introduction}

The suggestion by \pac\ (1986) that compact objects in the halo of
our Galaxy might be detected by means of gravitational microlensing
was confirmed in a rather spectacular fashion last September when the
EROS, MACHO, and OGLE collaborations each discovered their first
candidate microlensing events in the course of less than a month
(Alcock, \etal, 1993, Aubourg, \etal, 1993, and Udalski, \etal, 1993).
Since then, the total number of microlensing events discovered has
grown to about 60, and the implications of the microlensing results toward
both the Large Magellanic Cloud and the Galactic bulge remain somewhat
mysterious.

The basic physics of microlensing is quite simple.
If a compact object passes very close to the
line of sight to a background star, the light will be
deflected to produce two images of the star.
In the case of perfect alignment, the star will appear
as an `Einstein ring' with a radius of
$$
 r_E = \sqrt{ { 4 G M L x (1-x) \over c^2 } },
$$
where $M$ is the lens mass, $L$ is the observer-source distance
and $x$ is the ratio of the observer-lens and observer-source distances.
In a typical situation of imperfect alignment, the image appears
as two arcs.
On cosmological scales, many cases of quasars multiply imaged
by foreground galaxies are known ;
however, for stellar lensing the angular separation
of the lensed images is much smaller.
For a source distance of $50 \kpc \approx 10^{10} \au $ and
a lens distance of $10 \kpc$, the Einstein radius
is $r_E \approx 8 \sqrt{M / \msun} \au $ ; this
yields an angular separation of $\sim 0.001$ arcsecond which
is well below even the Space Telescope resolution, hence the term
`microlensing'.

However, in the point source approximation,
the lensing produces a net amplification of
the source by a factor
$$
  A = { u^2 + 2 \over u \sqrt{u^2 + 4} }
$$
 where $u = b / r_E$ and $b$ is the distance of the lens from
 the direct line of sight. (The above scale of $8 \au$ is much
larger than the radius of a star,
hence the point source approximation is accurate
for lens masses $\simgt 10^{-5} \msun$).
The amplification is approximately $u^{-1}$ for $u \simlt 0.5$, and
$1 + 2u^{-4}$ for $u \gg 1$, hence the amplification
 can be very large but falls rapidly for $u \simgt 1$.
Since objects in the Galaxy are in relative motion,
this amplification will be time-dependent; for a
typical lens transverse velocity of $200 \kms$, the duration is
$$
\hat{t} \equiv {2 r_E \over v_{\perp} }
  \approx 140 \sqrt{M / \msun} {\rm \; days}.
$$
This is a convenient timescale for astronomical observations,
and thus by sampling on a range of timescales the microlensing
searches may be sensitive to a wide mass range from small
planets of $\sim 10^{-6} \msun$ to black holes of $\sim 100 \msun$,
covering most of the plausible candidates.

\sec{Optical Depth}

The `optical depth' $\tau$ for gravitational microlensing is
defined as the probability that a given star is lensed with
$u < 1$ or $A > 1.34$ at any given time, and is
$$
  \tau = \pi \int_0^L {\rho(l) \over M} r_E^2(l) \, dl
$$
where $l$ is the distance along the line-of-sight and $\rho$ is
the dark matter density.
Since $r_E \propto \sqrt{M}$, while for a given $\rho$
the number density of lenses $\propto M^{-1}$,
the optical depth is independent of the individual MACHO masses.
Using the virial theorem, it is found that
$\tau \sim (v/c)^2$, where $v$ is the rotation speed of the Galaxy.
More detailed calculations (Griest, 1991) give an optical
depth for lensing by halo dark matter
of stars in the Large Magellanic Cloud (LMC) of
$$
  \tau_{\rm LMC} \approx 5 \times 10^{-7}
$$

This very low value is the main difficulty of the experiment;
only one star in two million will be amplified
by $A > 1.34$ at any given time, while the fraction of
variable stars is much higher, $\sim 3 \times 10^{-3}$.

\sec{Microlensing Signatures}

Fortunately, microlensing has many strong signatures which
can discriminate it from stellar variability :


\item{1)} Since the optical depth is so low, only one event should be
  seen in any given star.
\item{2)} The deflection of light is wavelength-independent,
   hence the star should not change color during the amplification.
\item{3)} The accelerations of galactic objects are negligible on timescales
of these events, hence
the events should be symmetrical in time, and have a shape
 derived from (2) with $u(t) = \sqrt{ u_{min}^2 + (v_{\perp} (t - t_{max}) /
r_E)^2 }. $
Examples of such light-curves are shown in Figure~2.

%
All these characteristics are distinct from known types of
intrinsic variable stars; most variable stars are periodic or
semi-regular, and do not remain constant for long durations.
They usually change temperature and hence color as they vary,
and they usually have asymmetrical lightcurves with a rapid rise and
slower fall.

In addition to these individual criteria,
if many candidate microlensing events are detected, there
are further statistical tests that can be applied:

\item{4)} The events should occur equally in stars of different colors
 and luminosities.
\item{5)} The distribution of impact parameter $u_{min}$ should
  be uniform from 0 to the experimental cutoff $u(A_{min})$.
\item{6)} The event timescales and peak amplifications should be uncorrelated.
(In practice these distributions will be modulated by the detection
efficiencies, which can be computed from simulations).

It is worth clarifying two common misconceptions: first,
this experiment
does {\it not} provide an estimate of $\Omega$, since
the amount of dark matter within 50 kpc of spiral galaxies is
approximately known from rotation curve data.
This represents
$\Omega \sim 0.05$: if $\Omega = 1$ either the halos must
extend far beyond 50 kpc or there must be intergalactic dark matter.
Secondly, the lenses are not required to be dark, merely significantly
fainter than the source stars. However, since the LMC is
located $30^o$ from the galactic plane, the optical depth
from known stars is only $\simlt 10^{-8}$ (Bahcall, \etal, 1994).

\sec{Microlensing Searches}

Due to the low optical depth, a very large number of stars must
be monitored over a long period to achieve a significant detection rate.
The optimal targets when searching for Machos in the galactic halo
are the Large and Small
Magellanic Clouds, the largest of the Milky Way's satellite galaxies,
since they have a high surface density of stars and are distant enough
at 50 and 60 kpc to provide a good path length through the dark halo.
This requires a Southern hemisphere observatory.

Another target for microlensing seaches is the Galactic bulge where a
higher microlensing optical depth is predicted due to lensing by
faint stars in the Galactic disk (Griest, \etal, 1991, \pac, 1991).
Naively, one might expect that microlensing toward the bulge would
not be of great relevance for the halo dark matter problem, but as
we shall see, this is not the case.

There are curently three microlensing surveys that are taking data and
have reported results: the EROS, MACHO, and OGLE collaborations. The
EROS collaboration operates two seperate experiments which observe the
LMC. They search for events lasting from days to months using photographic
plates taken at the ESO Schmidt and for events lasting less than a day
from a 40-cm telescope with a large format CCD array camera also located
at the ESO site in La Silla, Chile. They plan to upgrade to a dedicated 1-m
telescope with a very large format two color
CCD camera system (64 or 128 Mega-pixels)
in late 1995 or 1996. The OGLE collaboration now observes the Galactic
bulge with $\sim 75$ nights per year on the 1-m Swope telescope at
Las Campanas, Chile, but they also plan to upgrade to a dedicated
1.3-m telescope in 1995 or 1996.

The MACHO collaboration has full-time use of the
1.27-m telescope at Mt.Stromlo Observatory near Canberra, Australia, which
is used to image $0.5$ square degrees of sky simultaneously in two color
bands. The two foci are equipped with large
CCD cameras (Stubbs, \etal, 1993), each containing 4 Loral CCD chips of
$2048 \times 2048$ pixels. Most of the observing time is divided between
the LMC and bulge, with a small amount of time being spent on the SMC.

\sec{Microlensing Towards the LMC}

The first Galactic microlensing events every discovered were announced
simultaneously by the EROS and MACHO collaboration in the fall of 1993
(Alcock, \etal, 1993, Aubourg, \etal, 1993). These data initiated
a flurry of speculation as to whether the the halo might be entirely
baryonic or whether some non-halo population might be responsible for the
microlensing toward the LMC. Although the EROS and MACHO collaborations
have not yet announced definitive results on the microlensing optical depth
toward the LMC, preliminary results given in a number of conferences suggest
that the optical depth is both significantly below the prediction
($\tau \approx 5\times 10^{-7}$) for a ``standard halo" composed entirely
of Machos and significantly above the maximum possible value ($\tau < 10^{-8}$)
due to known
distributions of stars (Bahcall \etal, 1994). The possibility that these
events are not microlensing but rather a new variable star phenomena now
seems rather unlikely as spectral follow-up has confirmed that they appear
to be normal stars (Della Valle, 1994, Beaulieu, \etal, 1994,
Giraud, 1994, and Alcock, \etal, 1994a).

One possibility that remains viable is that the microlensing may be due to
faint stars in the LMC (Sahu, 1994), but this seems unlikely for several
reasons. First, to get the required event rate, one need to assume that
the mass of the LMC is almost entirely baryonic while the Milky Way is
apparently dominated by non-baryonic matter. It is difficult to see how
this could come about. Also, although microlensing by a self-consistent
model of the LMC which is massive enough to explain the observed events
has never been investigated, naive arguments suggest that the timescales of
the observed events could most easily be explained by microlensing by
objects of substellar mass.

Thus, the most reasonable conclusion is that the microlensing is due to
brown dwarfs in a halo or perhaps spheroidal population
(Giudice, Mollerach, \& Roulet, 1993), but that
the these objects do not provide the dominant mass for a ``standard halo\rlap."
It is tempting to conclude that there must be another component that dominates
the halo, except that, as we shall see, the microlensing optical depth toward
the galactic bulge is much higher than expected. Thus, another viable
possibility is that the dynamics of the inner galaxy are dominated by a massive
disk and bulge, and that Machos are the dominant component of a much smaller
than ``standard" halo.

\vbox{\hfil{
\epsfxsize=3.5in
\epsffile{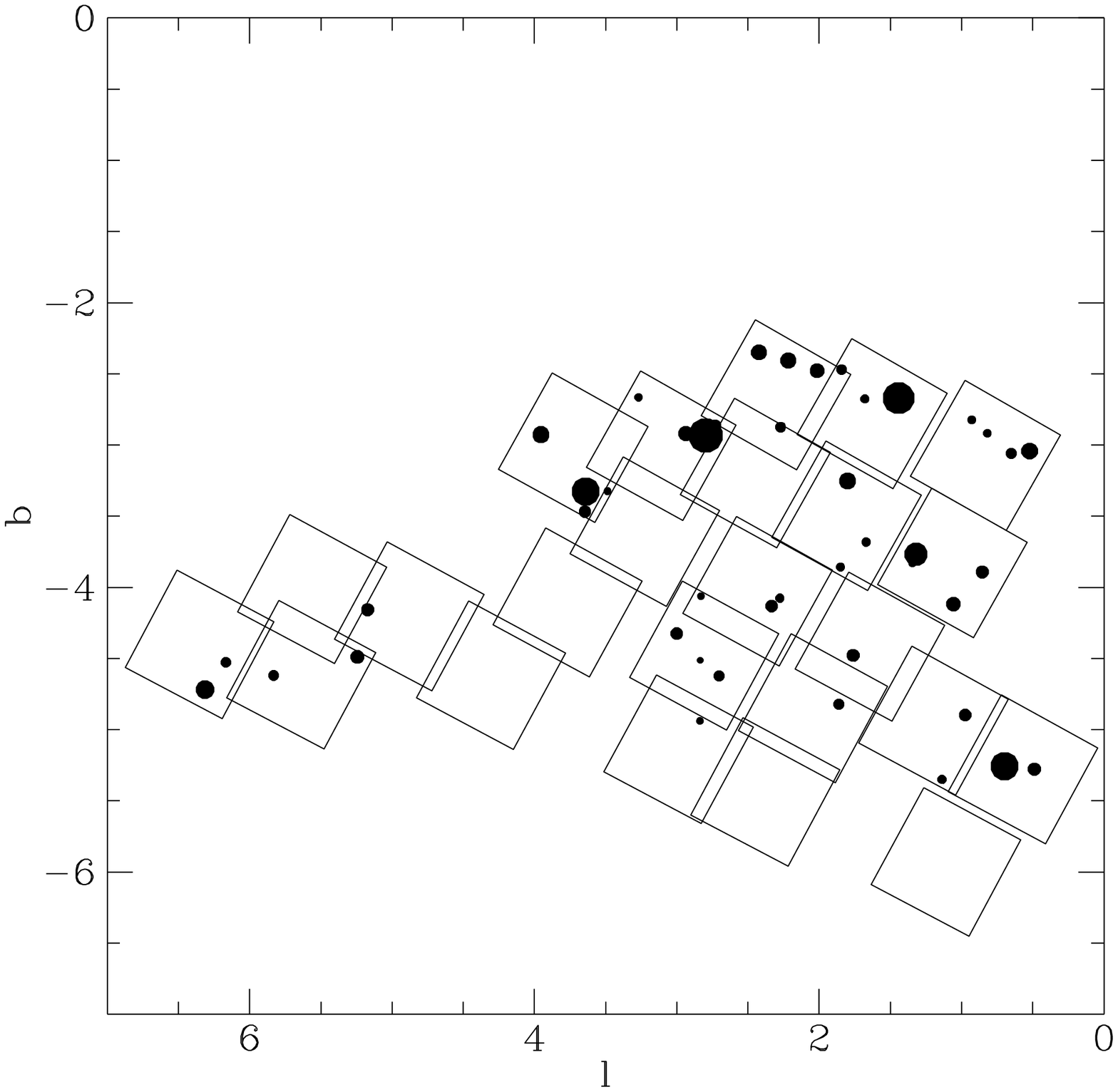}
\hfil}}
{\captpar \noindent
{\bf Figure 1.} The location of the 24 bulge fields that were observed
by the MACHO group and analyzed for this paper. The spots indicate the
locations of detected events, and the spot area is proportional to the
$\that$ value for the event.}
\vskip 0.1in

\sec{Microlensing Events Toward the Galactic Bulge}

Within a week after the EROS and MACHO teams announced the first microlensing
events detected toward the LMC, the OGLE collaboration had discovered the
first microlensing event ever seen toward the Galactic bulge (Udalski, \etal,
1993). Subsequent analyses by the OGLE group revealed about 10 additional
events, and revealed a very high optical depth towards Baade's window
(Udalski, \etal, 1994a). These results are fully consistent with the
MACHO results that will be discussed in the remainder of this paper.

\vbox{{
\epsfxsize=4.3in
\epsffile[18 164 593 718]{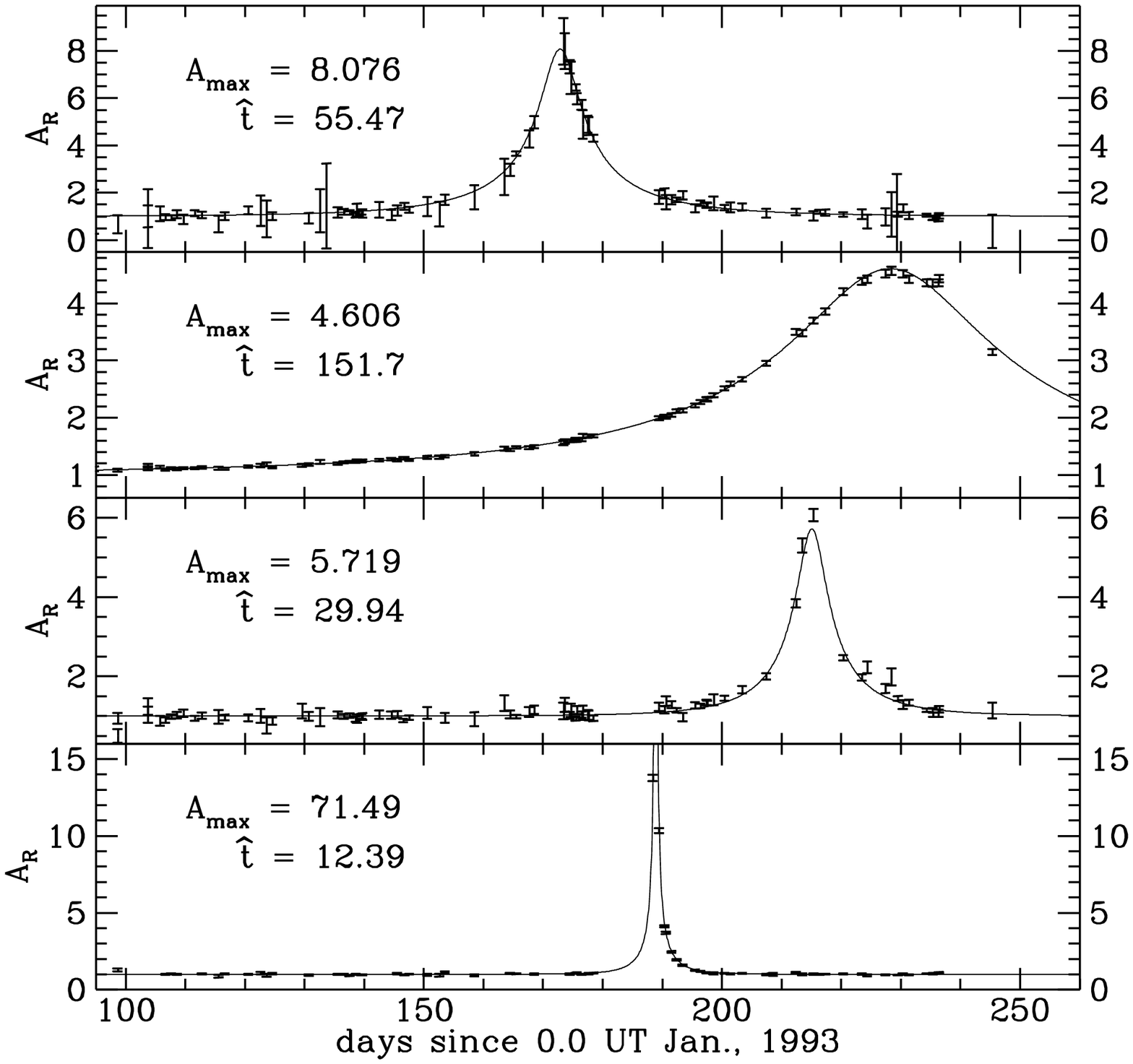}
\hfil}}
{\captpar \noindent
{\bf Figure 2.} MACHO-Red band lightcurves of 4 events from field 101
located at $\ell \approx 3.8$, $b\approx -3.1$.}
\vskip 0.1in

We have recently analyzed the 2-color light curve data for about 10 million in
24 fields toward the Galactic bulge. The location of these fields in
galactic coordinates (as well as the location of the events) in shown in
Fig. 1. Before starting the microlensing search, measurements with questionable
PSF chi square, crowding, missing pixel, or cosmic ray flags are flagged as
bad measurements and removed from further consideration. Then, we run matched
filters through the data in both colors, and 37,000 stars passed the
filter trigger threshold and were fit jointly in both colors
with a microlensing light curve.
These stars form our "level 1" candidate list and are subject to further
cuts on the difference between the $\chi^2$ values for the microlensing
and flux$=$const. fit, and the microlensing fit $\chi^2$ per d.o.f. Further
cuts are made on the number of data points near the peak and the average
estimated error. Four of the 45 candidates which pass these cuts are shown in
Fig. 2. A few of the 45 events may be variable stars, but the majority
are almost certainly microlensing events. A color magnitude diagram
showing the locations of these 45 stars as well as 7\% of the stars
within 2.5' of these stars is shown in Fig. 3.

\vbox{\hfil{
\epsfxsize=4.5in
\epsffile{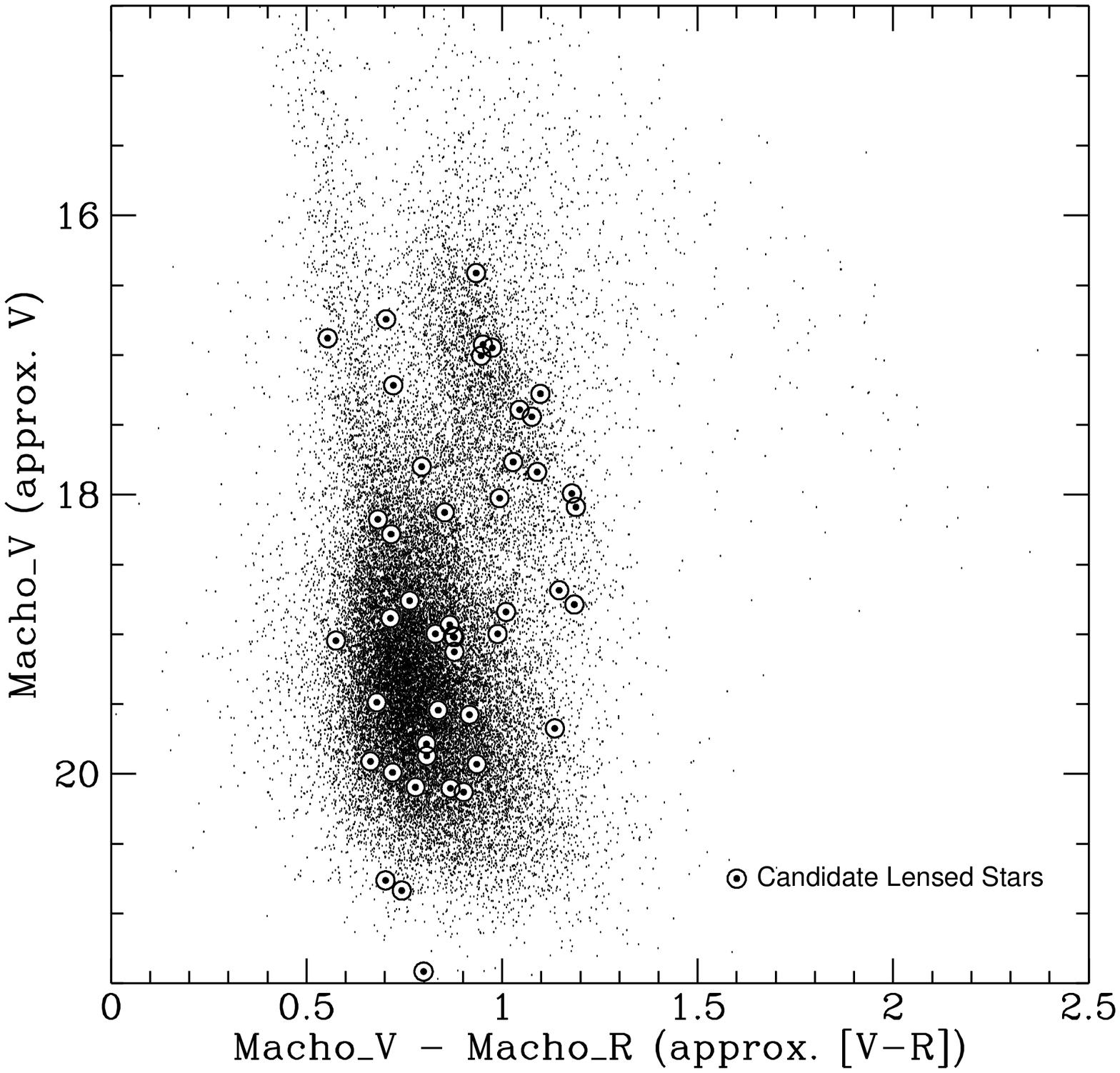}
\hfil}}
{\captpar \noindent
{\bf Figure 3.} A color magnitude diagram for the 45 microlensing
candidates and 7\% of the stars that fall within 2.5' of the candidates.}
\vskip 0.1in

\sec{Preliminary Optical Depth Estimates}

The primary measurable quantity in a gravitational microlensing experiment is
the microlensing optical depth which is defined to be the instantaneous
probability that a random star is magnified by a lensing object by more than
a factor of 1.34. This is a measure of the mass in microlensing objects
along the line of sight to the source stars. Experimentally, one can
define the measured optical depth as the amount of time that the
microlensed stars spend inside the Einstein ring (\ie are amplified by
more than 1.34).
$$ \tau_{\rm meas} = {1 \over E} {\pi\over 4}
                     \sum_i {\hat t_i \over \epsilon(\hat t_i)} \ . $$
where $E$ is the total exposure (in star-years), $\hat t_i$ is the
Einstein ring diameter crossing time, and $\epsilon({\hat t_i})$ is the
detection efficiency.

The detection efficiency $\epsilon$ is properly determined by adding
``fake" stars with the same luminosity function as the real stars to the
raw images. The brightnesses of these ``fake" stars can be modulated
according to randomly selected microlensing light curves, and then the
analysis can be run to see how many of these simulated events are recovered.
A much simpler procedure would be to add simulated microlensing light
curves to the light curve database, but this yields $\epsilon$ values that
are generally too large because it neglects the effects of systematic
errors in the photometry and the blending of stellar images which tends
to make lensing events more difficult to see. The $\epsilon$ values calculated
in this way are known as sampling efficiencies, and these are the only
efficiencies that we will present here. $\epsilon(\hat t)$ is plotted for
several different data cuts in Fig. 4.

\vbox{\hfil{
\epsfxsize=3.5in
\epsffile[18 194 593 527]{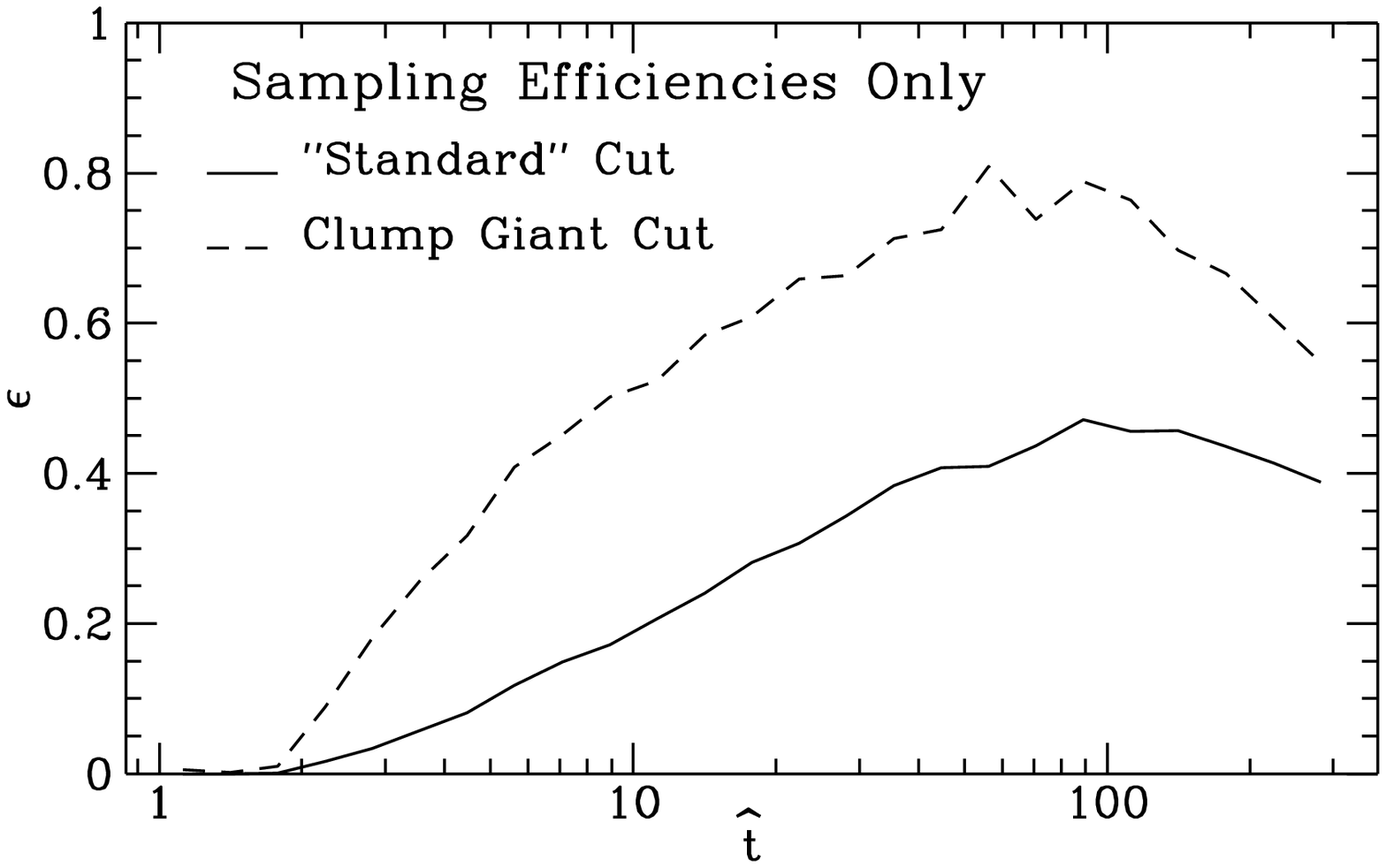}
\hfil}}
{\captpar \noindent
{\bf Figure 4.} Sampling Efficiencies are plotted for our standard and
clump giant cuts. The efficiencies are higher for clump giants because they
are much brighter than the average star.}
\vskip 0.1in

\def\blt{$\vert b\vert < 3.1$}

\centerline{\bf Microlensing Optical Depth Towards}
\centerline{\bf the Galactic Bulge $\times 10^6$}
\vskip 0.3 truecm

\begintable
 & | \multispan{6}\tstrut\hfil\ Confidence Levels \hfil &
     \cr
cut  | N | $0.025$ | $0.05$ | $0.16$ | {\bf measured} | $0.84$  | $0.95$ |
$0.975$ \cr
standard | 43 | $1.02$| $1.11$| $1.29$| {\bf 1.58}| $1.93$| $2.16$| $2.28$\nr
clump giant | 13 | $1.25$| $1.48$| $2.06$| {\bf 3.03}| $4.47$| $5.46$|
$5.96$\nr
\blt | 18 | $1.81$| $2.11$| $2.72$| {\bf 3.79}| $5.27$| $6.26$| $6.83$\nr
clump giant | 9 | $2.04$| $2.57$| $3.91$| {\bf 6.32}| $10.27$| $13.07$| $14.48$
 \nr
\ w/ \blt | | | | | | | |
    \endtable
{\captpar \noindent
{\bf Table 1.} The opticcal depth $\tau$ is listed in units of $10^{-6}$ for
the different cuts discussed in the text. Note that the \blt\ cuts
have selection effects that have not
been included in the confidence level estimation.}
\vskip 0.1in

Another problem that interferes with the determination of the microlensing
optical depth toward the Galactic Bulge is that many of the stars that we
see when we look towards the galactic bulge are foreground disk stars. If we
do not correct for this, then our estimates of the optical depth $\tau$ will
be underestimates. One way to avoid both the problem of foreground stars and
the complications due to stellar blends is to concentrate on a class of stars
that is both bright and ``known" to be in the galactic bulge:
the ``clump giant"
stars. These are relatively low mass core helium burning giants--the
horizontal branch of a metal rich population. Their location is marked on
the color magnitude diagram in Fig. 3.

The measured optical depths for a number of different cuts on the data are
shown in Table 1. This table displays the measured optical depth and various
confidence level limits on the microlensing optical depth for several
different cuts on the bulge data sample. The optical depth $\tau$ corresponding
to a confidence level, $P$, is the
$\tau$ value such that a fraction $P$ of simulated data sets have a measured
optical depth $> \tau$. The simulated data sets are constructed
assuming Poisson statistics and an actual $\hat t$ distribution that is
identical to the observed $\hat t$ distribution.

The standard cut referred to in Table 1
is quite similar to the one which generated
the 45 events shown in Fig. 2, but a few more cuts had to be added for
consistency with the monte carlo events used for the efficiency estimation.
Two additional types of cuts are referred to in Table 1. The first is the
clump giant cut alluded to above. The fact that this cut seems to imply
a noticeably higher optical depth ($\tau = 3.0{+1.5\atop -0.9}\times 10^{-6}$)
than the standard cut ($\tau = 1.58{+0.35\atop -0.28}\times 10^{-6}$)
suggests that the full efficiency corrections and/or the foreground star
corrections are important for the full sample.

The other cut referred to in Table 1 is a cut to
include only the 5 fields which
have a central galactic latitude of $|b| < 3.1$. This cut is intended to
emphasize the contribution of microlensing by objects in the galactic disk
as opposed to the galactic bulge or bar. A galactic bar elongated along the
line of sight has been proposed as the source of the large microlensing optical
depth seen toward the galactic bulge by the
OGLE group (\pac, \etal, 1994)
and by Zhou, Spergel, and Rich (1994). We
(Alcock, {\it et.al.}, 1994b) and Gould (1994b) have
suggested that a heavy, ``maximal" disk might account for the large optical
depth seen toward the bulge. The results shown in Table 1 do seem to suggest
that microlensing by the disk is contributing a large fraction of the total
optical depth, but we should caution that the cut at $|b| < 3.1$ has been
selected after the distribution of events with $b$ was known. So, this cut
can be considered to have been made in a biased way. A more sophisticated
analysis is now under way to determine if, in fact, these data indicate that
the ``maximal disk" model is preferred.

\sec{An ``Alert" Event}

Fig. 5
shows the light curve of the first gravitational microlensing event
ever detected substantially before its peak amplification. The time of
discovery is indicated. An IAU circular (IAUC 6068) was submitted about
24 hours after the event was discovered, and a great deal of photometric
and spectroscopic data was collected during the course of the event. The
team of Benetti, Pasquini, West, and de Lapparent obtained spectra on
6 different nights from the NTT and 3.6m telescopes at ESO, and have
confirmed that the spectral features of the star did not change as the
star rose to a peak amplification of $A=3.5$ and then dropped back down to
its usual brightness (IAUC 6069 and 6071). Similar data was obtained at
a number of large telescopes including a spectrum obtained
by Dick Hunstead at the AAT on Sept. 2. The ESO and AAT data indicate
that star is classified as a K0
class II-IV star with a radial velocity of $\sim -200$km/sec.

\vbox{\hfil{
\epsfxsize=4.0in
\epsffile[18 174 593 527]{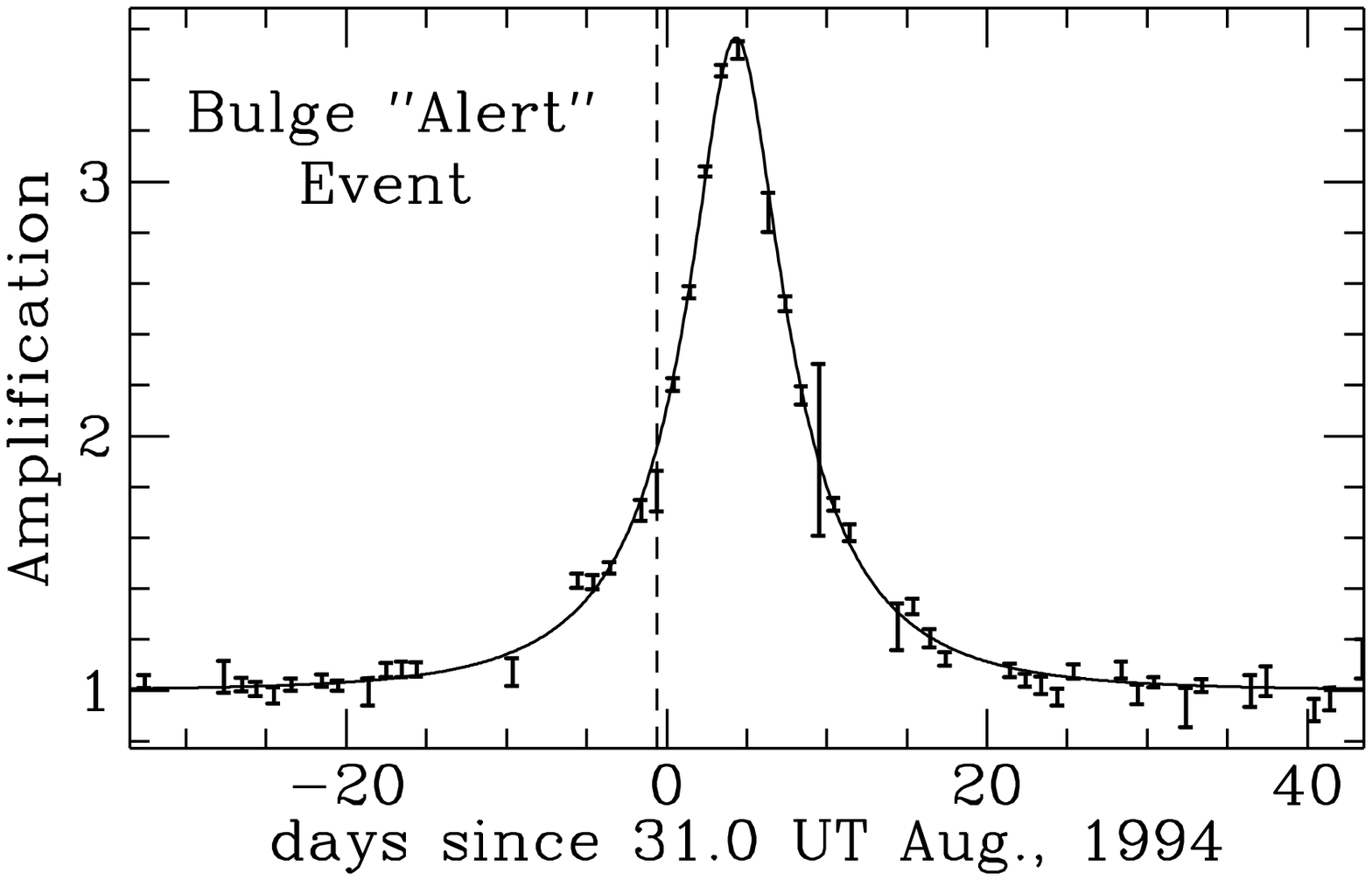}
\hfil}}
{\captpar \noindent
{\bf Figure 5.} The light curve of the first microlensing event discovered
substantially before peak amplification. The vertical dashed line is the
time of discovery.}

\vbox{\hfil{
\epsfxsize=4.0in
\epsffile[18 174 593 527]{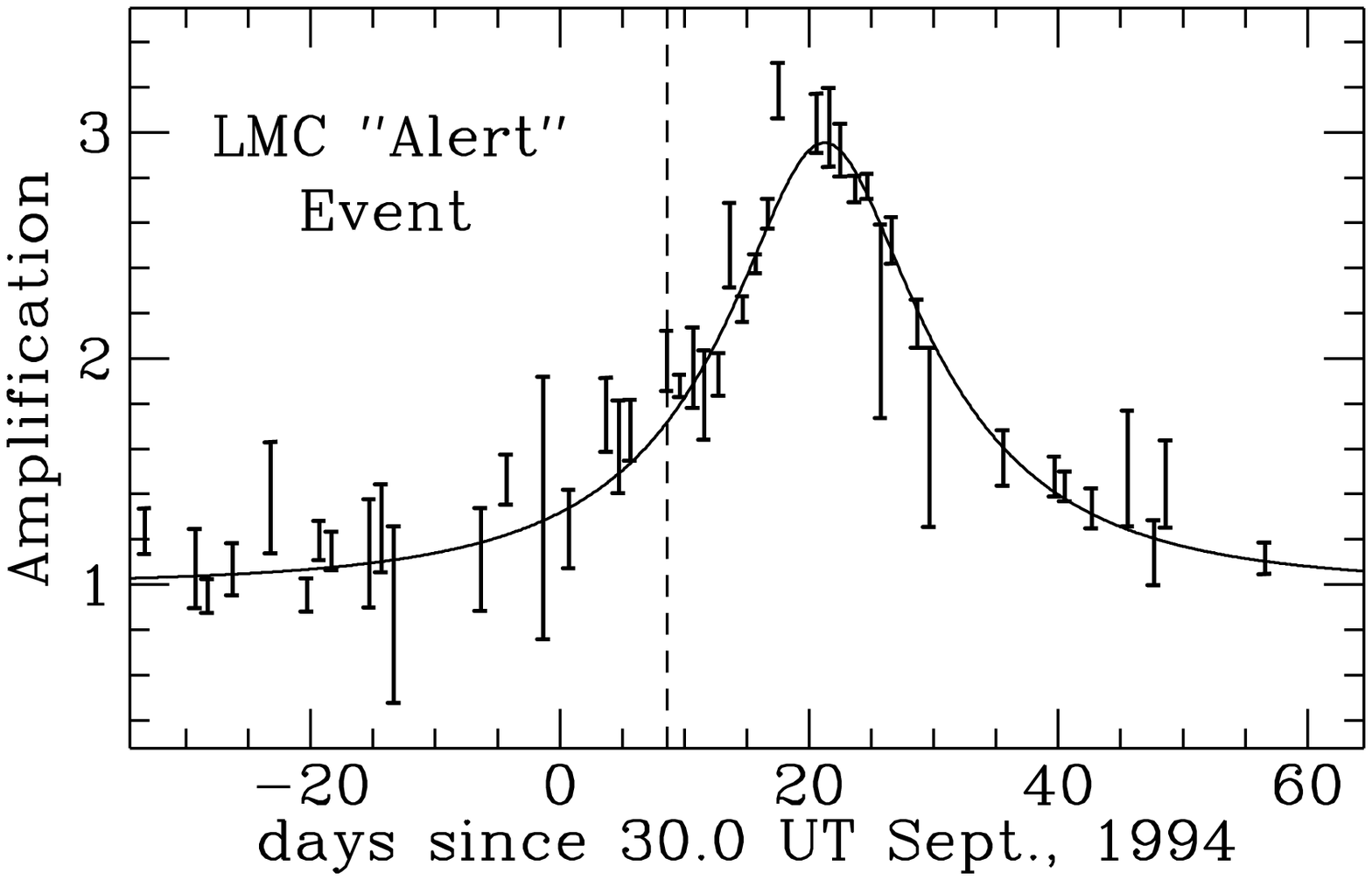}
\hfil}}
{\captpar \noindent
{\bf Figure 6.} The light curve of the first microlensing event
discovered in progress towards the LMC. The vertical dashed line is the
time of discovery.}
\vskip 0.1in

An alert system has also been demonstrated by the OGLE collaboration
(Udalski, {\it et.al.}, 1994c) who have detected two moderate
amplification events in progress.

At the time of this conference, the first microlensing event ever detected
in real time toward the LMC appears to be in progress!! The light curve of
this event is shown in
Fig. 6.
Giraud (private communication) obtained a spectrum of this star
near peak amplifiction, and confirmed that it is a normal F star.

\vfil\break

\sec{A Parallax Event}
\vbox{\hfil{
\epsfxsize=4.5in
\epsffile[18 184 593 500]{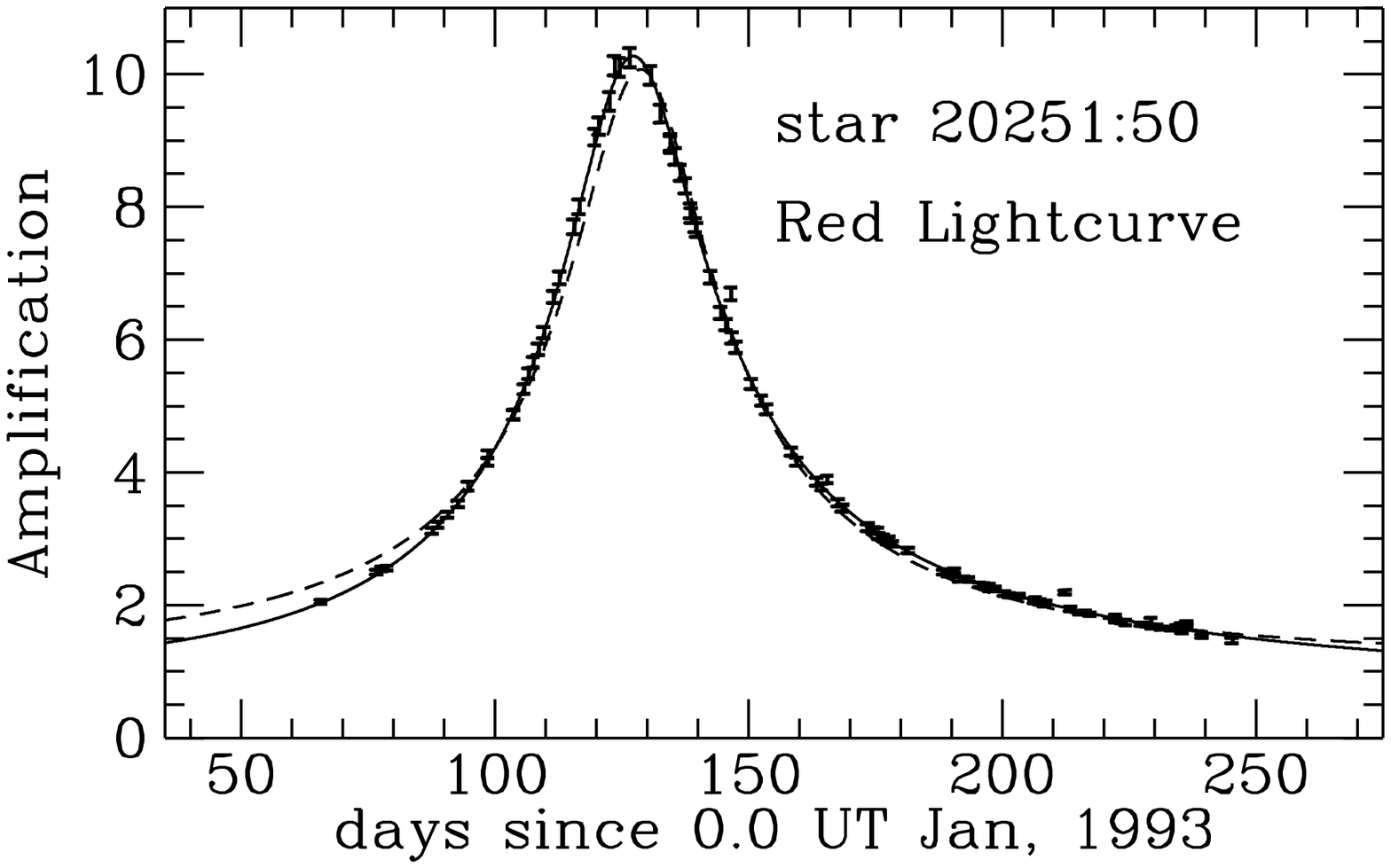}
\hfil}}
{\captpar \noindent
{\bf Figure 7.} The MACHO-Red band light curve for the longest event
yet detected by the MACHO project. The solid curve is the fit light curve
which includes the effect of the earth's motion, while the dashed curve
is the best fit ignoring the motion of the earth.}
\vskip 0.05in

Fig. 7 shows the light curve of the longest timescale microlensing event
that we have detected, and the first to show a significant asymmetry in
it's light curve. This asymmetry can be explained as a parallax
effect--the change in the earth's velocity over the hundreds of days that
this event has lasted has caused a deviation from the simple 3-parameter
light curve that describes most microlensing events. This light curve is
well fit by a model which includes the orbital motion of the earth with
the transverse velocity of the microlensing object projecting to a
transverse velocity of 54 km/sec at the position of the earth. This gives
us the information we need to determine the mass of the microlensing object
as a function of the earth-lens distance. The implied mass is shown in
Fig. 8. It seems clear from Fig. 8 that the lensing object must very likely
be a low mass star in the Galactic disk or a brown dwarf in the galactic
bulge.

\vbox{\hfil{
\epsfxsize=3.0in
\epsffile[18 184 593 510]{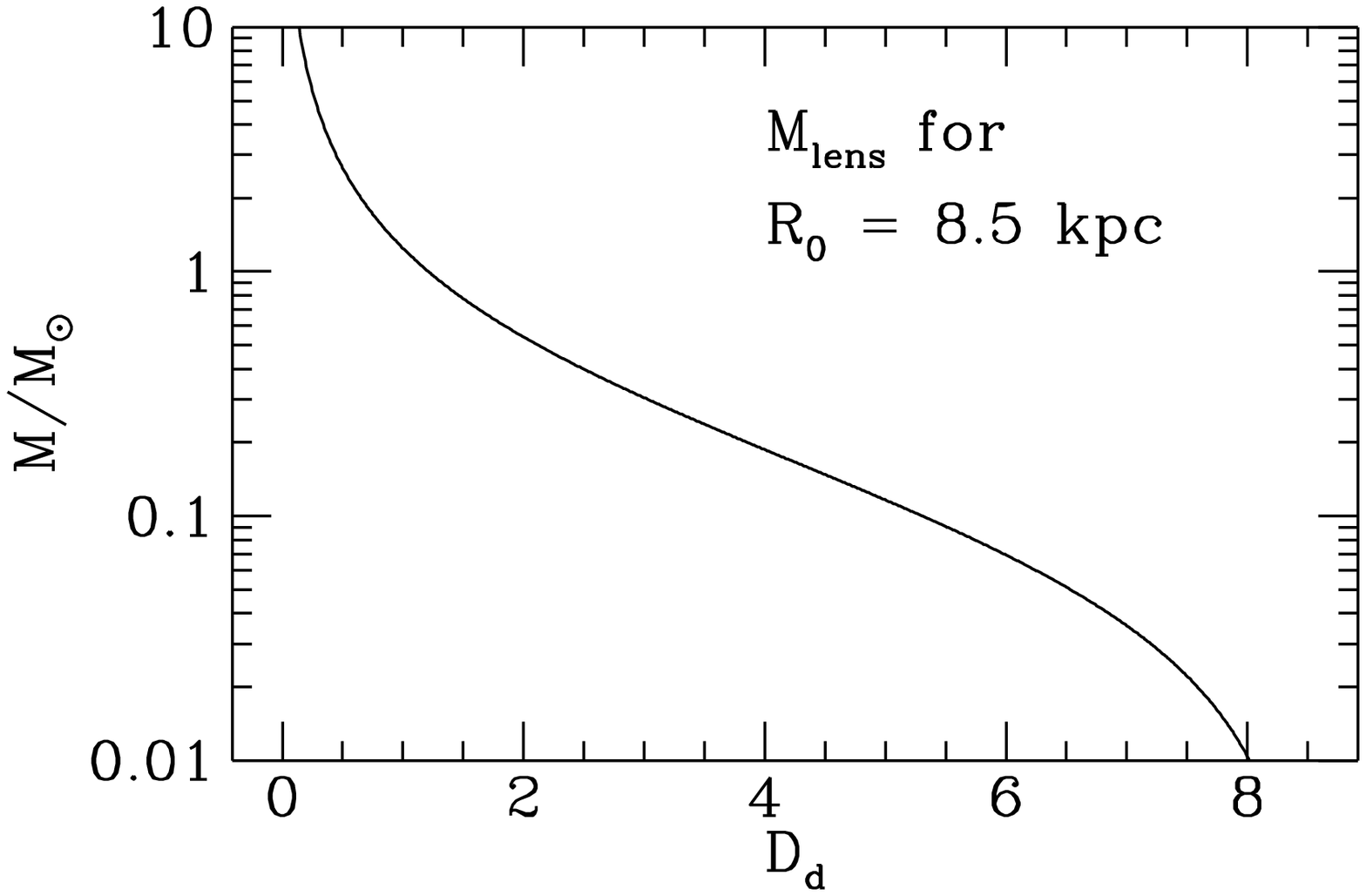}
\hfil}}
{\captpar \noindent
{\bf Figure 8.} The implied mass is plotted as a function of the distance
to the lens for the parallax event.}
\vskip 0.05in

Much progress could be made toward resolving the mysteries of microlensing
if the parallax effect could be observed for a large number of microlensing
events. However, it is only for the very long events that the earth will
move enough during the event for such an effect to be detectable. This
situation could be remedied if a small satellite could observe events from
a solar orbit, as suggested by Gould (1994a-b). The satellite
would need to be alerted to events detected in progress from the ground, but
this capability has already been demonstrated by the OGLE and MACHO groups.

\sec{MACHO Observations of Microlensing by a Binary}

\vbox{\hfil{
\epsfxsize=4.5in
\epsffile[18 174 593 527]{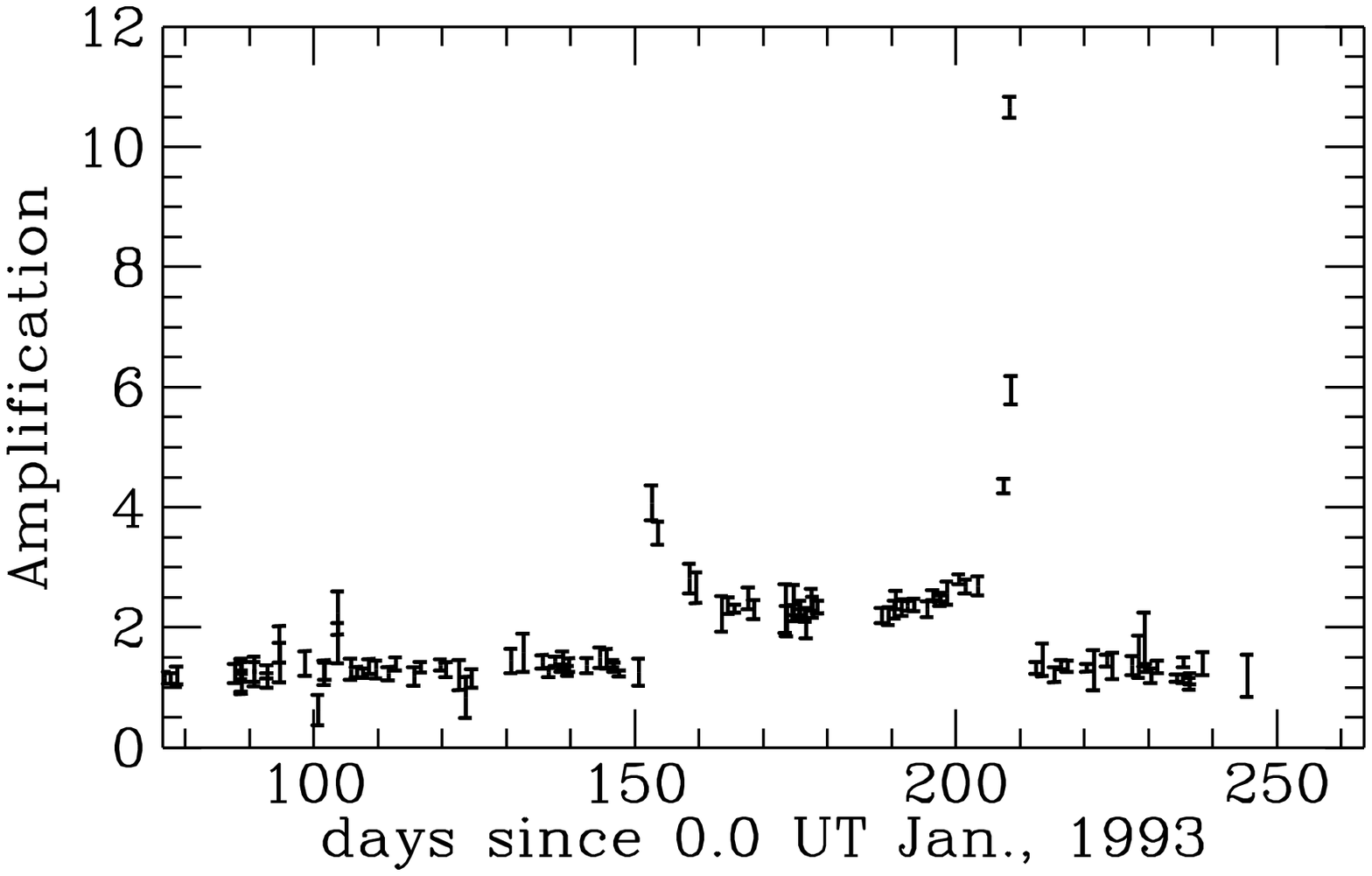}
\hfil}}
{\captpar \noindent
{\bf Figure 9.} The MACHO-Blue band light curve of the binary lens event
first seen by OGLE.}
\vskip 0.1in

Fig. 9 shows light curves of a microlensing event that clearly shows
characteristics of a binary lens with ``caustic crossings" where the
source star crosses into (or out of) a region where two extra images
are created (making a total of 5). This event was first discovered by the
OGLE collaboration (Udalski, {\it et.al.}, 1994c), but the
MACHO data provides a nice confirmation with very good coverage of
the second caustic crossing which was largely missed by OGLE. In fact,
the MACHO coverage of the second caustic crossing is so good, that the
finite angular size of the source star is resolved in the MACHO data.
The divergent amplification of the 4th and 5th images near a caustic
can be shown to have the form $A \approx d^{-1/2}$ where $d$ is the
distance to the caustic on the interior side. The modification of this
formula for a finite size source star with a realistic limb darkening
law has been derived by Schneider and Wagoner (1987). Fig. 10
shows a fit using the 4 MACHO Blue band data points closest to the
caustic using a 3 parameter model with the Schneider and Wagoner profile.
The fit $\chi^2 = 0.009$ for 1 d.o.f. This fit fixes the time for the
caustic to cross the diameter of the star to be 10 hours. Using the
OGLE fit value for the angle of the caustic crossing, we find that the
time for the lens position to cross the diameter of the star is about
8 hours. This implies a lens velocity of
$v_t = x(48 {\rm km/sec}) R_{\rm star}/R_\odot$, where $x$ is the
fractional distance between us and the lensing object. This will
lead to a contraint on the mass of and distance to the lens.

\vbox{\hfil{
\epsfxsize=3.5in
\epsffile[18 174 593 527]{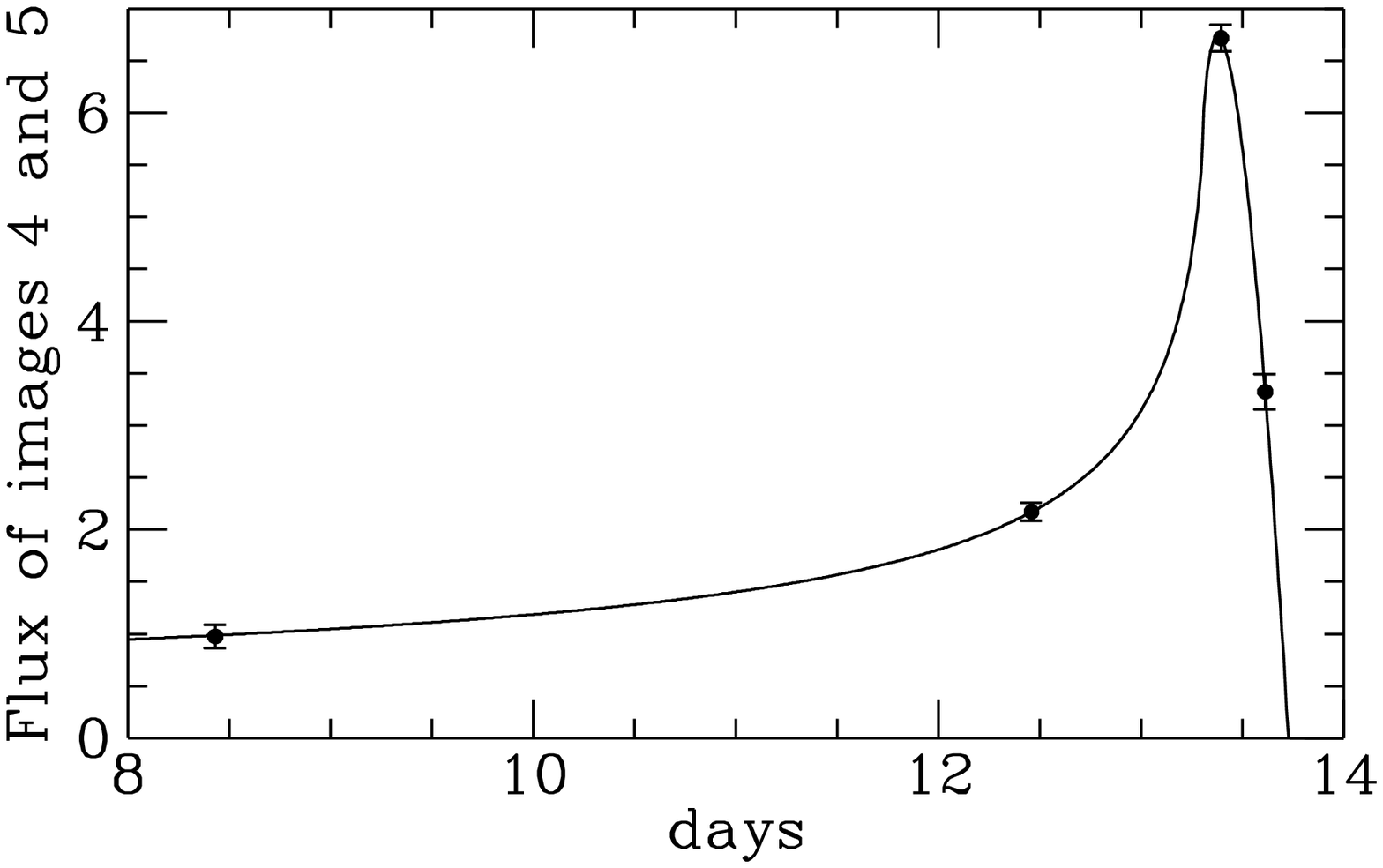}
\hfil}}
{\captpar \noindent
{\bf Figure 10.} A fit to the 2nd caustic crossing of the binary lens event
using the MACHO-Blue band data.}
\vskip 0.1in

\sec{4. Conclusions}

The microlensing surveys toward the galactic bulge have yielded a number of
surprizes including exotic events such as the binary and parallax events.
Although they now have the status of curiosities that are interesting in
their own right, they also have the virtue that we know more about the
properties of the lenses in these events than the lenses in the other
microlensing events. With the help of the recently developed ``Alert" systems,
we will soon be obtaining higher quality data on events that are discovered
in progress which should increase the chances of detecting these exotic events.
With higher quality data
coupled with more observations covering a larger range of $\ell$ and $b$, we
hope to shed some light on the major bulge microlensing mystery, which is:
Why is the optical depth so high? The resolution of this puzzle will
probably help to determine whether the Galactic halo is made predominantly of
baryonic or non-baryonic matter.

\sec{ACKNOWLEDGEMENTS}
This research was supported by the National Science Foundation through the
Center for Particle Astrophysics at UC, Berkeley and by the Department of
Energy through the LDRD program at the Lawrence Livermore National Laboratory.

\sec{REFERENCES}

\def\pac{Paczy{\'n}ski}
\def\apj{{ \it ApJ}}

\hi{Alcock, C., Akerlof, C.W., Allsman, R.A., Axelrod, T.S., Bennett, D.P.,
Chan, S., Cook, K.H., Freeman, K.C., Griest, K., Marshall, S.L., Park, H.-S.,
Perlmutter, S., Peterson, B.A., Pratt, M.R., Quinn, P.J., Rodgers, A.W.,
Stubbs, C.W., \& Sutherland, W., 1993, {\it Nature}, {\bf 365}, 621.}

\hi{Alcock, C., Allsman, R.A., Axelrod, T.S., Bennett, D.P.,
Cook, K.H., Freeman, K.C., Griest, K., Marshall, S.L.,
Perlmutter, S., Peterson, B.A., Pratt, M.R., Quinn, P.J., Rodgers, A.W.,
Stubbs, C.W., \& Sutherland, W., 1994a, in preparation.}

\hi{Alcock, C., Allsman, R.A., Axelrod, T.S., Bennett, D.P.,
Cook, K.H., Freeman, K.C., Griest, K., Marshall, S.L.,
Perlmutter, S., Peterson, B.A., Pratt, M.R., Quinn, P.J., Rodgers, A.W.,
Stubbs, C.W., \& Sutherland, W., 1994b, {\it ApJ}, in press.}

\hi{Aubourg, E., Bareyre, P., Brehin, S., Gros, M., Lachieze-Rey, M.,
Laurent, B., Lesquoy, E., Magneville, C., Milsztajn, A., Moscosco, L.,
Queinnec, F., Rich, J., Spiro, M., Vigroux, L., Zylberajch, S., Ansari, R.,
Cavalier, F., Moniez, M., Beaulieu, J.-P., Ferlet, R., Grison, Ph.,
Vidal-Madjar, A., Guibert, J., Moreau, O., Tajahmady, F., Maurice, E.,
Prevot, L., \& Gry, C., 1993, {\it Nature}, {\bf 365}, 623}

\hi{Bahcall, J.N. Flynn, C., Gould, A., \& Kirhakos, S., 1994, {\it Nature},
in press.}

\hi{Beaulieu J.P., Ferlet R., Grison Ph., Vidal-Madjar A., Kneib JP.,
 Maurice E., Prevot L., Gry C., Guibert J., Moreau O., Tajahmady F.,
 Aubourg E., Bareyre P., Brehin S., Gros M., Lachieze-Rey M., Laurent B.,
 Lesquoy E., Magneville C., Milsztajn A., Moscoso L., Queinnec F., Rich J.,
 Spiro M., Vigroux L., Zylberajch S., Ansari R., Cavalier F., \& Moniez M.,
 1994, preprint.}

\hi{Della Valle, M., 1994 {\it A. \& A.}, {\bf 287}, L31}

\hi{Giudice, G.F., Mollerach, S., \& Roulet, E., 1993, (preprint:
CERN-TH.7127/93)}

\hi{Gould, A., 1994a, {\it ApJ}, {\bf 421}, L75}

\hi{Gould, A., 1994b, preprints.}

\hi{Griest, K., 1991, \apj, {\bf 366}, 412}

\hi{Griest, K., Alcock, C., Axelrod, T.S., Bennett, D.P.,
Cook, K.H., Freeman, K.C., Park, H.-S.,
Perlmutter, S., Peterson, B.A., Quinn, P.J., Rodgers, A.W.,
\& Stubbs, C.W.,
1991, {\it ApJ Lett.}, {\bf 372}, L79}

\hi{Kiraga M., and \pac, B. 1994, {\it ApJ Lett.}, in press.}

\hi{\pac, B, 1986, \apj, {\bf 304}, 1}

\hi{\pac, B. 1991, {\it ApJ Lett.}, {\bf 371}, L63}

\hi{\pac, B., Stanek, K.Z., Udalski, A., Szymanski, M., Kaluzny, J.,
    Kubiak, M., Mateo, M., \& Krzeminski W., 1994 {\it ApJ Lett.},
    {\bf 435}, L113}

\hi{Sahu, K.C., 1994, {\it Nature}, {\bf 370}, 275.}

\hi{Schneider, P., \& Wagoner, R.V., 1987, {\it ApJ}, {\bf 314}, 154}

\hi{Stubbs, C., Marshall, S.L., Cook, K.H., Hills, R., Noonan, J.,
Akerlof, C.W., Axelrod, T.S., Bennett, D.P., Dagley, K., Freeman, K.C.,
Griest, K., Park, H.-S., Perlmutter, S., Peterson, B.A., Quinn, P.J.,
Rodgers, A.W., Sosin, C., \& Sutherland, W.,
1993, {\it SPIE Proceedings}, {\bf 1900}, 192}

\hi{Udalski, A., Szymanski, M., Kaluzny, J., Kubiak, M., Krzeminski, W.,
Mateo, M., Preston, G.W., \& Paczynski, B., 1993, {\it Acta Astronomica},
{\bf 43}, 289}

\hi{Udalski, A., Szymanski, M., Stanek, K.Z., Kaluzny, J., Kubiak, M.,
Mateo, M., Krzeminski, W., Paczynski, B., \& Venkat, R., 1994a,
{\it Acta Astronomica}, {\bf 44}, 165}

\hi{Udalski, A., Szymanski, M., Kaluzny, J., Kubiak, M.,
Mateo, M., Krzeminski, W., Paczynski, B., \& Venkat, R., 1994b,
{\it Acta Astronomica}, {\bf 44}, 227}

\hi{Udalski, A., Szymanski, M., S. Mao, R. Di Stefano, J.
Kaluzny, M. Kubiak, M. Mateo, \& W. Krzeminski, 1994b, {\it ApJ Lett.},
in press.}

\hi{Zhou, H.S., Spergel, D.N. and Rich, R.M., 1994, preprint.}


\vfill
\bye
%
\centerline{\LARGE \bf ACKNOWLEDGEMENTS}

\end